\documentclass [12pt]{article}
\usepackage{graphics}
\usepackage{hangcaption}
\renewcommand{\baselinestretch}{2}

\title{Rapidity Gap of Weakly Coupled Leptoquark Production in 
$ e p$ Collider}

\author{ T. K. Kuo and Taekoon Lee 
        \\
        \\
        Department of Physics\\
        Purdue University\\
        West Lafayette, IN 47907}

\date{}
\begin{document}   
\maketitle
\begin{abstract}
We show that the  current bounds on the leptoquark couplings imply that if 
leptoquarks are produced in $e p $ collisions, a significant fraction of 
them could  form a leptoquark-quark bound state. The decay of the 
bound state has a distinct event shape with rapidity gap. A possible 
application of this observation in the leptoquark search at HERA is
discussed. 

\end{abstract} 
 
\def\thepage{PURD-TH-97-03}
\thispagestyle{myheadings}
\newpage
\renewcommand{\baselinestretch}{2}
\pagenumbering{arabic}
\addtocounter{page}{0}
\newcommand{\be}{\begin{equation}}
\newcommand{\ee}{\end{equation}}
\newcommand{\bear}{\begin{eqnarray}}
\newcommand{\eear}{\end{eqnarray}}

There has been a considerable interest in the possible existence of 
leptoquarks. However, the couplings to quarks and leptons of leptoquarks
 which are  accessible in
the lepton-proton collider are severely bounded by various low energy physics.
Generally the couplings from these bounds are very weak,
of order electromagnetic  interaction, or weaker \cite{buch1,davison1}.
Because of this the lifetime of a light leptoquark is in the 
QCD hadronization time scale, and thus  leptoquarks
produced in $e p$ collisions could form a leptoquark-quark bound state before
their decay. Because the leptoquark-quark bound state is a color singlet, its
eventual decay should leave  a distinct  event shape which might be 
characterized by the presence of  rapidity gap \cite{bj}. 
This rapidity gap could provide a unique tool in the search
of leptoquarks in $ep$ collisions. To our knowledge, this idea has not been
explored in the search of leptoquarks \cite{zeus,h1_95}.

To be specific, we consider a scalar leptoquark with non-zero
 fermion number, with
coupling to the left-handed quarks and leptons only. Choosing this specific
leptoquark is for the sake of simplicity only, and applying  our discussion
 to
 other leptoquarks is straightforward. Then the most general 
effective lagrangian that preserves baryon and lepton number conservation and
the standard model symmetry is given by \cite{buch2}
\be
{\cal L}= \lambda\, \Phi\, \bar{q}_{L}^{c} i \sigma_{2} l_{L} + h.c. 
 \label{1}
\ee
where $q_{L}, l_{L}$ denote the left-handed quark and lepton doublets.

The coupling $\lambda$ is bounded by the quark-lepton universality 
\cite{buch1}:
\be
 \lambda \le \frac{m_{LQ}}{1.7 \mbox{TeV}}
 \label{2}
\ee
which corresponds to $\lambda \le 0.1$ for the leptoquark mass
 $m_{LQ}=200 \, \mbox{GeV}$.
This bound assumes that leptoquarks are the only deviation to the standard
model. Without the assumption the bound could be weakened.

The decay width of the leptoquark from the effective lagrangian in 
eq. (\ref{1}) is
\be
\Gamma_{\Phi}= \frac{\lambda^2}{8 \pi} m_{LQ}.
 \label{3} \ee
For $\lambda=0.1, m_{LQ}=200 \, \mbox{GeV}$ we have a very narrow width
\be
\Gamma_{\Phi} \approx  80 \,\mbox{MeV}.
\ee 
Indeed this narrow decay width is a generic feature of leptoquarks accessible
at collider \cite{buch2}. Also note that the decay width is 
small even when compared
 to the
leptoquark hadronization scale. In terms of
QCD hadronization-time scale the leptoquark is quasi-stable.  
This observation is the central basis of the discussion
in this letter.

Leptoquarks with very narrow decay width could be produced asymptotically
in the $s$-channel in $ep$ collisions. The cross section for leptoquark
 production
from the effective lagrangian in eq. (\ref{1}) is
\be
\sigma= \frac{\pi}{4 s} \lambda^2 f_{q}(\frac{m_{LQ}^{2}}{s})
\label{5} 
\ee
where $f_{q}(x)$ is the quark distribution function in proton, and $\sqrt{s}$
is the c.m. energy. 

\begin{figure}[ht]
\leavevmode
\begin{center}
\rotatebox{90}{
\resizebox{8cm}{10cm}{%
\includegraphics* [5cm,5cm][18cm,22cm]{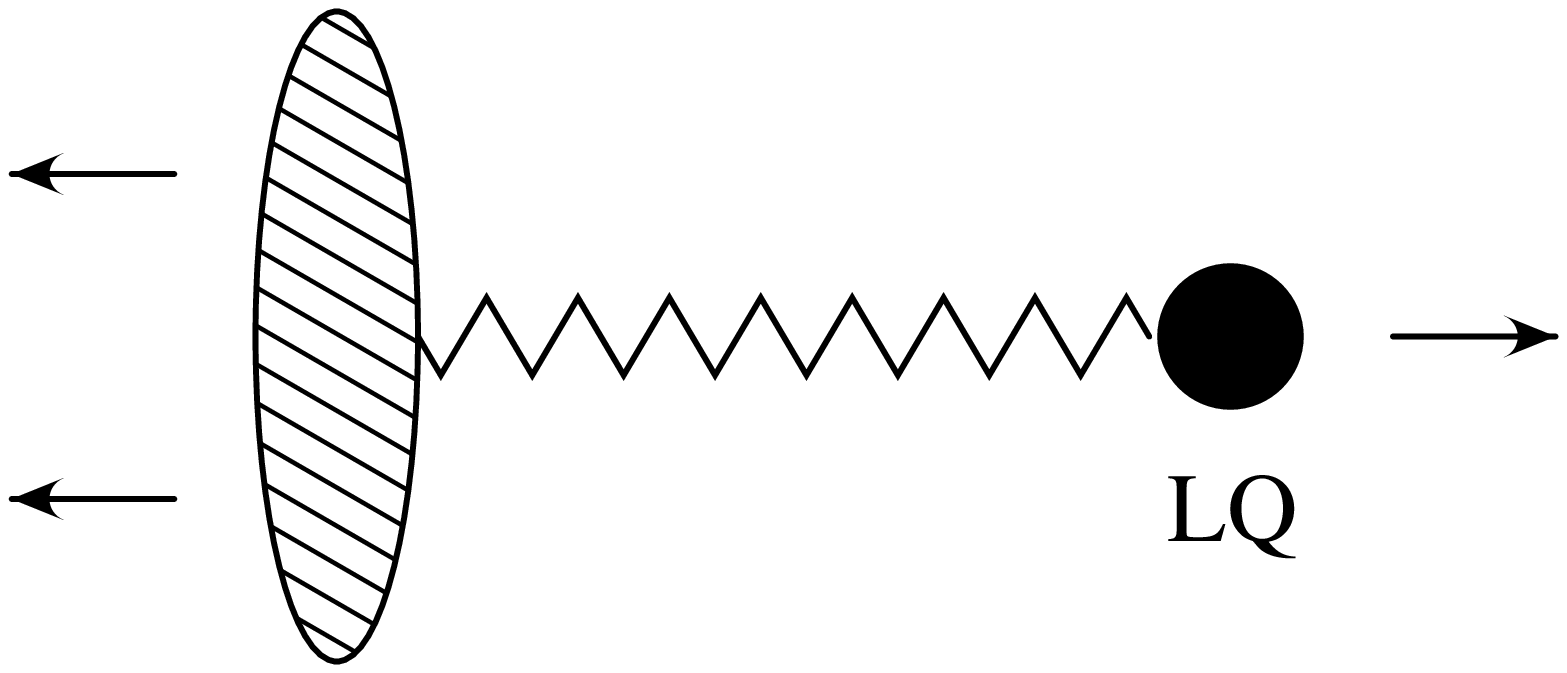}} } 
\end{center} 
\isucaption{Schematic of leptoquark production in the c.m. frame.
The shaded blob denotes proton remnant, and the zigzag line denotes color
electric  string. }
\end{figure}

The $s$-channel production of leptoquark combined with its long lifetime 
leads to rapidity gap in the final state. To see this, consider the 
lepton-proton collision in the c.m. frame. An $s$-channel production
of a leptoquark would leave the leptoquark on the beam axis ( See Fig.1 ).
 Since the leptoquark carries 
color, there would be a color electric string attached between the leptoquark
and the outgoing proton remnant. This color string will eventually decay into
hadrons. In the leptoquark rest frame, the hadronization of the color string
would occur at the leptoquark end first, and then propagates toward the
proton remnant. This picture is also true in the c.m. frame if we ignore the
relativistic effect due to the small leptoquark velocity in the c.m. frame,
 which is
\be
\frac{v}{c} \approx 0.37 
\ee 
for $m_{LQ}=200 \, \mbox{GeV}$ at the HERA energy $\sqrt{s}= 296 \, \mbox{GeV}$.
Thus in the c.m. frame, the leptoquark forms a leptoquark-quark bound state,
denoted by $\Psi_{\Phi q}$, at the early stage of the color string
hadronization, and decouples from the rest of the hadronization process.
The bound state $\Psi_{\Phi q}$ will eventually decay into a lepton and
a jet. The appearance of a jet in the decay of the bound state is easy
to understand. At the decay of the bound state, an energetic lepton and 
a quark will be emitted from the center of the bound state 
in a back-to-back
 direction, and the  primary quark from the center 
 would drag the spectator
valence quark along its motion, forming a jet.

The invariant mass of
the jet can be easily estimated to be
\be
M_{\mbox{jet}} = \sqrt{m_{LQ} ( m_{\Psi_{\Phi q}} - m_{LQ})}.
\ee 
The mass difference between the bound state and leptoquark is
\be
 m_{\Psi_{\Phi q}} - m_{LQ} \approx  m_{\mbox{B}} - m_{\mbox{b}} 
\approx 1 \, \mbox{GeV} 
\ee 
For $m_{LQ}=200 \, \mbox{GeV}$
\be
M_{\mbox{jet}} \approx 14 \, \mbox{GeV}. 
\ee 

Since $\Psi_{\Phi q}$ is color singlet, a rapidity gap should appear
between the jet from the decay of the bound state and the proton remnant on
 the 
forward direction. One might worry about the contamination of the 
rapidity gap from the hadronization of the color string.
 However, the contamination from the
hadronization of the color string should be minimal. This is because
the $s$-channel production of leptoquark would confine the color string
on the beam axis, and hence the hadrons from the string decay should move
along the beam axis, leaving a clean gap between the jet and the proton
remnant.

The cross section for $\Psi_{\Phi q}$ production  can be estimated by
 counting the
number of the leptoquarks that survive beyond the leptoquark 
hadronization-time $T_{0}$. Here we assume that all leptoquarks surviving
beyond $T_{0}$ form bound states. For the leptoquark  hadronization-time 
we assume that the average radius of the valence quark to the
center of a heavy meson is  a reasonable estimation of $T_{0}$. 
From the B-meson,
we have \cite{hwang}
\be
T_{0} \approx ( 400\, \mbox{MeV} )^{-1}.
\ee

Using the decay width and cross section in eqs. (\ref{3}), (\ref{5}),
 we can readily derive
a relation
\be
\frac{N(\lambda)}{N(\lambda_{0})} = \left( \frac{\lambda}{\lambda_{0}}\right)
^{2} e^{ 1- \left( \frac{\lambda}{\lambda_{0}}\right)^{2}}
 \label{11} \ee
where $N(\lambda)$ is the number of leptoquarks surviving beyond $T_{0}$ at
a given luminosity, and $\lambda_{0}$ is defined by
\be
\Gamma_{\Phi}(\lambda_{0})\, T_{0} =1.
\ee 
For $m_{LQ}=200 \, \mbox{GeV}$
\be
\lambda_{0}=0.22
\ee 
and at  HERA energy $\sqrt{s}=296 \, \mbox{GeV}$,
\be
N(\lambda_{0})= L \cdot \frac{\pi}{4 s} \lambda_{0}^{2}
 f_{q}(\frac{m_{LQ}^{2}}{s}) e^{-1}= 2000 \cdot \frac{L}{100 \mbox{pb}^{-1}},
\ee 
where $L$ is the luminosity.

\begin{figure}[ht]
\leavevmode
\begin{center}
\rotatebox{-90}{
\resizebox{10cm}{12cm}{%
\includegraphics* [3cm,4cm][18cm,22cm]{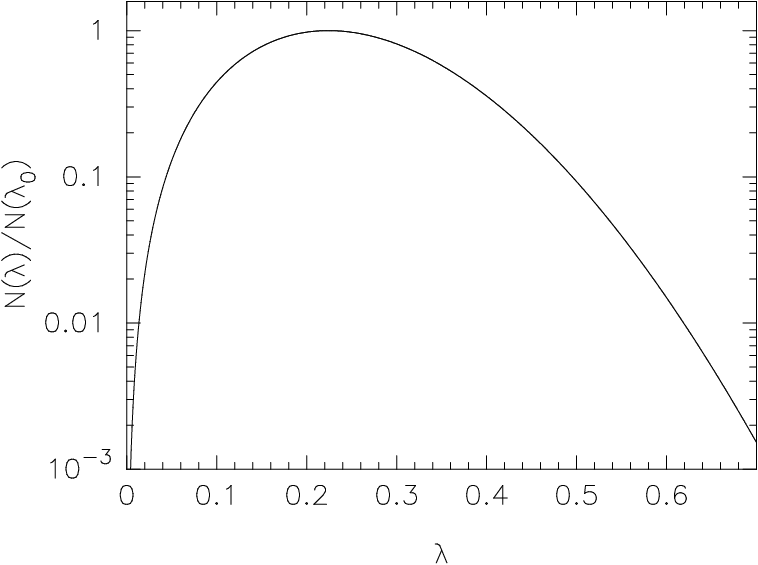}} } 
\end{center} 
\isucaption{
Relative number of leptoquarks that survive beyond the leptoquark
hadronization time. }
\end{figure}

A plot of the function in eq. (\ref{11}) is shown in Fig. 2. With the HERA
luminosities, an  observable
number of  $\Psi_{\Phi q}$ can be produced with the coupling in the range
\be
0.01 \le \lambda \le 0.6.
\ee 
Thus the rapidity gap in the final state jet could be an effective tool
in the search of leptoquarks over a wide range of couplings.

The main background for the leptoquark production is the $t$-channel 
lepton-proton deep inelastic scattering (DIS). However, jets from DIS
are not expected to have large rapidity gap because the hadrons from the
 decay of
the color string  between the struck quark and proton remnant would fill
the rapidity space between the jet and  proton remnant \cite{agis,zeus2}. 
The detailed effect on
the events with rapidity gap from DIS  should 
be reliably estimated using Monte Carlo simulation.

In conclusion,  we have presented an intuitive picture of leptoquark
hadronization that leads to event shape with rapidity gap. This rapidity
gap in the final state jet could be exploited to enhance the signal to 
background ratio in the search of leptoquarks in HERA. The background from
DIS  for the signal with rapidity gap could be estimated with  Monte Carlo
 simulation.
We encourage a detailed Monte Carlo study incorporating the idea discussed 
here. 

\vspace{0.6in}
\noindent
We are grateful to V. Barnes for useful discussion.

 \end{document}